\documentclass[printer]{tMOP2e}

\usepackage{graphicx}
\usepackage{amsmath,amssymb}
\newcommand{\ie}{i.e.~}
\renewcommand{\mod}[1]{\mathrm{mod}\;#1}
\newcommand{\proj}[1]{\pi_{#1}}
\newcommand{\hilb}{\mathcal{H}}
\newcommand{\ket}[1]{|#1\rangle}
\newcommand{\bra}[1]{\langle#1|}
\newcommand{\ctrled}[1]{\mathrm{c-}\!#1}

\newcommand{\rmi}{{\rm i}}
\newcommand{\rme}{{\rm e}}
\newcommand{\rmd}{{\rm d}}
\newcommand{\rmpartial}{{\rm\partial}}
\newcommand{\fl}{}

\begin{document}

\doi{10.1080/0950034YYxxxxxxxx}
\issn{1362-3044}
\issnp{0950-0340} %\jvol{00}

\markboth{Ko\v s\'\i k, Miszczak, Bu\v zek}{Quantum Parrondo's game}

\title{\itshape Quantum Parrondo's game with random strategies\footnote{We
dedicate this paper to Sir Peter Knight on the occasion of his 60th birthday.} }

\author{
J. Ko\v s\'\i k\thanks{$^\dagger$Email: kosik@quniverse.org}$^\dagger$\\[6pt]
Research Centre for Quantum Information, Slovak Academy of
Sciences\\D\'ubravsk\'a cesta 9, 84511 Bratislava, Slovakia\\[10pt]
J. A. Miszczak\thanks{$^\ddagger$Email: miszczak@iitis.gliwice.pl}$^\ddagger$\\[6pt]
Institute of Theoretical and Applied Informatics, Polish Academy of Sciences\\
Baltycka 5, 44-100 Gliwice, Poland\\[10pt]
V. Bu\v zek\thanks{$^\S$Email: buzek@savba.sk}$^\S$\\[6pt]
Research Centre for Quantum Information, Slovak Academy of Sciences\\
D\'ubravsk\'a cesta 9, 84511 Bratislava, Slovakia\\[6pt]
and \\[6pt]
Quniverse, L{\'\i}\v{s}\v{c}ie \'{u}dolie 116, 841 04 Bratislava, Slovakia\\[6pt]
\received{v1.0, released April 2007}}

\maketitle

\begin{abstract}
We present a quantum implementation of Parrondo's game with
randomly switched strategies using 1) a quantum walk as a source
of ``randomness'' and 2) a completely positive (CP) map as a randomized evolution. The
game exhibits the same paradox as in the classical setting where
a combination of two losing strategies might result in a winning strategy.
We show that the CP-map scheme leads to
significantly lower net gain than the quantum-walk scheme.
\end{abstract}

%\pacs{03.67.–a, 05.40.Fb, 02.50.Le}

\section{Introduction}
The theory of games \cite{NM44a} studies models in which several parties try to maximize their gains by selecting different
strategies that are allowed by the rules of a particular game. This theory can be applied in many different areas such as
resolutions of economical or political conflicts, investigations in an evolutionary biology, psychology,
etc. In the field of computer science the game theory is used to model distributed or
parallel computing.

Games are formalized by assuming that all parties
can choose from  a set of well-defined strategies, and that a
deterministic payoff function is defined for any choice of strategies. In
the classical game theory a
strategy is considered to be a  state of some specific physical system, which may
interact with other systems (strategies) according to  a given prescription (a set of rules associated with the game).
If strategies are associated with states of a physical system then it is natural to ask
what would happen if this system obeys laws of quantum
physics. This brings us to a notion of \textsl{quantum games}, where
strategies of each party are quantum states and manipulations with
strategies  are described by completely positive (CP) maps. The payoff function is then a quantum
observable on the tensor product of state spaces of all parties.
A nontrivial aspect of quantum games is the possibility of
a superposition of strategies, which may significantly affect the expected
payoff. At this point it should be noted that there is no canonical quantization procedure
of classical games. Quantum games are games with specific rules that include for instance a possibility
to consider superposition of strategies.

Among first models of quantum games that have been extensively studied is the so-called \textsl{Prisoner's dilemma} \cite{EWL99a}.
In the ``classical'' version of the game, two
suspects (prisoners), denoted as Alice and Bob, are tried by a prosecutor who offers
each of them separately to be pardoned
if they provide evidence against the other suspect. Now both suspects
may choose either to cooperate, \ie to not to comply with the request of the prosecutor , or to defect.
Different combinations of behavior lead to a payoff shown in
\ref{tab:prisoner-dilemma}. The optimal strategy for both suspects is to
cooperate; however this selection of strategies is unstable in the sense that
any player can separately improve his/her payoff, if the other player does not
change his/her strategy. On the other hand, the strategy $(D,D)$ is stable. It
has been shown \cite{Nas51a}, that the stable selection of strategies
(an equilibrium) exists under rather general conditions. In a quantum version of the
game, each player possess a qubit, whose state determines whether the player
will cooperate or defect. Both Alice and Bob entangle their qubits, then
separately (locally) apply unitary operators on their respective qubits, and then
disentangle the qubits. The measurement on both qubits yields the expected
payoff. It has been proven that if the entanglement between the qubits is
maximal, $(D,D)$ ceases to be stable; a new stable selection of strategies
emerges, which is also optimal.
% \begin{table}
% \label{tab:prisoner-dilemma}
% \caption{The payoff table for the Prisoner's dilemma game for each player (Alice, Bob)
% either cooperating (C) or defecting (D).}
% \begin{indented}
% \item[]\begin{tabular}{@{}lll}
% \br
% &Bob: C&Bob: D\\
% \mr
% Alice: C&(3,3)&(0,5)\\
% Alice: D&(5,0)&(1,1)\\
% \br
% \end{tabular}
% \end{indented}
% \end{table}

\begin{table}
\label{tab:prisoner-dilemma}
\caption{The payoff table for the Prisoner's dilemma game for each player (Alice, Bob)
either cooperating (C) or defecting (D).}
\begin{tabular}{@{}lll}
\toprule
&Bob: C&Bob: D\\
\colrule
Alice: C&(3,3)&(0,5)\\
Alice: D&(5,0)&(1,1)\\
\botrule
\end{tabular}
\end{table}

In Ref.~\cite{Mey99a} the author discussed the ``penny-flip'' model, in which two
players take turns applying their strategies; the payoff is computed after a
(short) sequence of turns. It has been proven that one of the player has an
optimal strategy (a definitive advantage) over the other one, provided he uses
quantum operations, while the other uses stochastic operations.
Moreover, it turns out that a two-person zero-sum game does not need to have an
equilibrium, when both players use quantum operations on their strategy spaces.

Sir Peter Knight and his collaborators have recently analyzed various aspects of quantum walks (for more details
see Refs.~\cite{Sanders2003,Knight2003a,Knight2003b,Knight2004,Carneiro2005}). In particular, they have
investigated physical implementations of quantum walks. In the present paper we will present a quantum implementation of Parrondo's game with
randomly switched strategies using  quantum walks as a source
of ``randomness''.
We will also analyze a situation when completely positive (CP) maps are used as randomized evolutions. We will show
that the game exhibits the same paradox as in the classical setting where
a combination of two losing strategies might result in a winning strategy.
Our paper is organized as follows: In Sec.~2 we will briefly describe a classical Parrondo's game, in Sec.~3
we will show how to implement a random choice of strategies using quantum walks. Numerical simulations of quantum
Parrondo's game will be presented in Sec.~4.
In Sec.~5 we will implement random choice of strategies using general completely positive maps and corresponding
numerical simulations will be presented in Sec.~6. Finally, in Sec.~7 we will analyze connections between the three versions
of Parrondo's game  discussed in the paper.

\section{Parrondo's game - an overview}
Parrondo's game \cite{HA99a,PHA00a} is a 1-player paradoxical game (the player plays
``against the environment''). The player repeatedly chooses from among two strategies
$A$,$B$. Each strategy involves a coin flip; the player adds or subtracts one
unit to his capital depending on the flip outcome. The coin is biased, and the
bias may depend on the amount of capital accumulated so far. We may choose the
bias of both coins to be such that if sequences of strategies $AA\dots A$ or
$BB\dots B$ are played then, the capital converges to $-\infty$. However, if we switch
between the strategies, the capital may converge to $+\infty$.

We restate the above arguments in a rigorous way:
\begin{definition}
{(\bf Parrondo's game,\cite{PHA00a})}
\label{def:parrondo}
Parrondo's game is a sequence $\{s(n)\in\{A,B\}:n\in\mathbb{N}\}$ where $A,B$ are
two strategies. Both strategies consist of a coin toss and adding or subtracting
one unit of capital to the player's account according to the result of the toss.
The probability that $A$ wins is $p$; the probability that $B$ wins is $p_0$ if
the capital is multiple of 3, and $p_1$ otherwise.
\end{definition}
We see that Parrondo's game is characterized by three coefficients
$p,p_0,p_1$, which determine the bias of both coins and the overall evolution of
the capital. The capital of the game $c(n)$ is a random variable of the number
of coin tosses $n$. If its mean value $\langle c(n)\rangle$ increases
(decreases), the game is called winning (losing). If $s(n)=A$ for all $n$ and
$p=\frac{1}{2}-\epsilon$, then the game is obviously losing. If $s(n)=B$, the
conditions for $p_0,p_1$ can be derived from the properties of the stationary
distribution of the Markov process $q(n)= c(n)\;\mathrm{mod}\;3$ (for more details see
Refs.~\cite{PD04a,MB02a}). It turns out that this sequence of strategies is losing iff
\begin{equation}
\label{eq:1}
  p_0<\frac{1-2p_1+p_1^2}{1-2p_1+2p_1^2}\;.
\end{equation}
Parrondo's paradox rests in the fact that some sequences of strategies can
nevertheless be winning. One such example is the sequence $\{s(n)\}=AABB\cdots$
(the strategy $A$ is used if $n\equiv k(\mathrm{mod}\;4),k=0,1$, and $B$ is used
otherwise) or random mixture of strategies, when $A$ or $B$ is played at each
step with probability $\frac{1}{2}$ \cite{PD04a,MB02a}. This is true, for
example, for $p_0=\frac{1}{10}-\epsilon,p_1=\frac{3}{4}-\epsilon$.
\begin{figure}
  \label{fig:1}
  \centering
  \includegraphics[width=.8\textwidth,keepaspectratio=true]{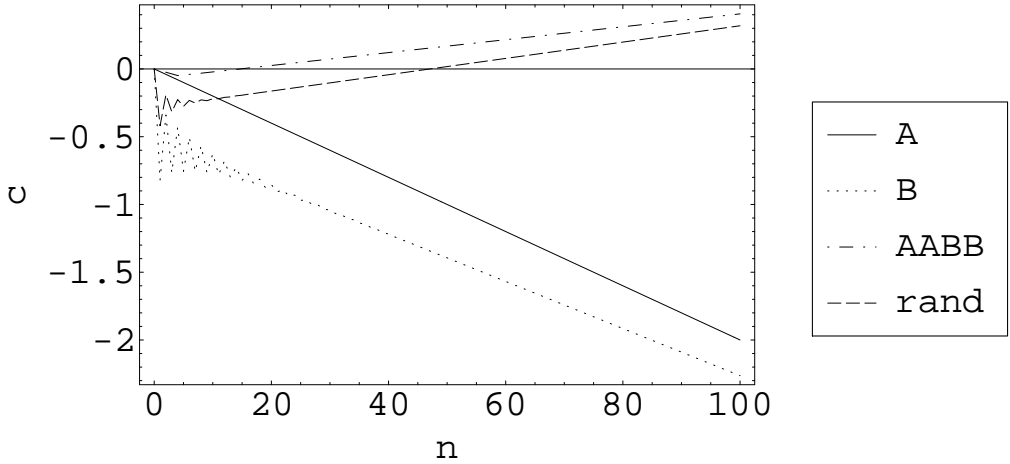}
  \caption{The expected capital $c(n)$ versus the number of steps $n$ for the
  Parrondo's game with the sequence of strategies
  $A\cdots$ (solid line),$B\cdots$ (dotted line),$AABB\cdots$ (slashed-dotted line) and random choice (slashed line) of strategies.
  The biases of the coins are
  $p=\frac{1}{2}-\epsilon$, while for the coin $B$ is
  $p_0=\frac{1}{10}-\epsilon,p_1=\frac{3}{4}-\epsilon$,
  ($\epsilon=\frac{1}{100}$). The initial capital is equal to 0.}
\end{figure}
The time dependence of the expected capital is shown in Fig.~1.

Parrondo's game may be thought of as a stochastic motion of a particle on the
line \cite{PD04a}.
For example (see Ref.~\cite{TAM03a}), any game driven by one coin which depends on the
amount of  capital modulo $L$ may be thought of as a stochastic motion on the
line is governed by the master equation
\begin{equation}
\label{eq:master}
P_x(n+1) = p_{x-1}P_{x-1}(n) + q_{x+1}P_{x+1}(n) \, ,
\end{equation}
where $P_x(n)$ is the probability that the capital amounts to $x$ after $n$ coin
tosses,  $p_{x}$ probability of the winning coin toss when capital is equal to
$x$, and $q_x=1-p_x$
Eq. (\ref{eq:master}) is just
the discretization of the Fokker-Planck equation
\begin{equation}
\label{eq:fokkerplanck}
\frac{\rmpartial P(\xi,t)}{\rmpartial t} = -\frac{\rmpartial}{\rmpartial
\xi}\left[F(\xi)P(\xi,t)\right]+\frac{1}{2}\frac{\rmpartial^2}{\rmpartial \xi^2}P(\xi,t) \, ,
\end{equation}
where $F(\xi)$ is the drift coefficient.
The discrete version of $F(\xi)$ is $F_x = p_x - q_x$
and the Parrondo's game is equivalent to the diffusion of a particle in the
potential
\begin{equation}
V_x = -\frac{1}{2}\sum_{y=1}^x\ln\left( \frac{p_{x-1}}{1-p_x} \right)  \, .
\end{equation}

The mean position of the particle is equivalent to the
expected capital. An application of the strategy is equivalent to turning on some
potential, which will cause the particle to drift in a certain direction. The
potential corresponding to the strategy $A$ is linear, while
the capital (=position) dependence of the strategy $B$ is modelled by a sawtooth
potential with a period equal to 3. By periodic switching of the potential on and off, the
particle can drift in either direction. This is an example of a Brownian motor,
when a thermal movement of the particle is directed by means of an external
source with global (overall) zero effect.

\section{Random choice of strategies with quantum walk}
Quantum games which have properties of Parrondo's game were
proposed in Refs.~\cite{MG05a,FNA02a}.
In Ref.~\cite{MG05a} the authors considered a scheme which is essentially equivalent
to our quantum walk scheme (see below), except that they do not use  a qubit
which ``randomly'' determines which strategy we use ($\ket{d}$ in our notation).
Hence, they
are constrained to deterministic  strategies sequences. Moreover, the state of
the  quantum coin which determines whether we win or lose one unit of the capital is
reset after each step. We decided to keep the state of the coin unchanged after
each step, possibly enforcing quantum interference effects. Our model may lead
to a higher rate of capital growth (see Fig.~\ref{fig:4}) depending on the
initial state of the coin which affects the ``random'' choice of strategies.

In Ref.~\cite{FNA02a} the authors considered the quantization of a classical stochastic motion
with a finite memory, which also leads to the Parrondo's effect. This was attained
by keeping the state of $n$ last ``coin tosses'' in a quantum register and  using
a sequence of unitary operators acting on one qubit of the register depending on
the state of other qubits in the register.
In what follows
we will focus, on the ``quantization'' of a random sequence $AB\dots$. If the
condition in Eq.~(\ref{eq:1}) is satisfied, this game is winning.

There is no unique way how to quantize the Parrondo's game. We should require
that
the amount of capital be encoded in the state of a quantum register with base
states from $\hilb=\{\ket{x}: x\in\mathbb{Z}\}$. Classically, updating of the
capital can be achieved by a random walk conditioned by the coins (strategies)
$A,B$. The ``quantization'' of the random walk was performed  in
Ref.~\cite{ADZ93a} as  a controlled permutation on $\hilb$, with
an additional register holding the result of the coin toss (unitary operation).
The connection of this dynamics with a classical Markov process is shown in
Ref.~\cite{RSS+04a}: It may be thought of as a random walk in 1 dimension with
an arbitrary bias to move in either direction, which contains an additional
``interference'' term between left and right steps in order to preserve the unitarity.
For consistency, we can also use the quantum coin tosses for the simulation
of random choice of strategies. Since the strategy $B$ requires dependence of
the coin toss on the state of $\ket{x}$ modulo 3, we need an additional control qubit
$\ket{o}$ which determines whether $x$ is divisible by 3 or not. This register
can be reset after each application of the strategy, based on the information
stored in other registers.

A quantum walk \cite{ADZ93a} is a unitary evolution (of a particle, for
simplicity) similar to a discrete random walk. The state of the particle is a
vector from the Hilbert space $\hilb$, which is spanned by the edges of some
underlying oriented graph. We restrict ourselves to regular graphs.
\begin{definition}(Quantum walk in 1D)
Let
$\hilb_C=\mathrm{span}\{\ket{c}:c=0,1\}$ (the coin space),
$\hilb_X=\mathrm{span}\{\ket{x}:x\in\mathbb{Z}\}$ (the position space)
and $\hilb=\hilb_C\otimes\hilb_X$ be the Hilbert space of the quantum walk. Let
$T_0,T_1$ be operators on $\hilb_X$ such that
$T_0\ket{x}=\ket{x-1},T_1\ket{x}=\ket{x+1}$, $U\in SU(2)$ and
$\proj{j}=\ket{j}\bra{j},\;j=0,1$ be projection operators on $\hilb_C$.
Then the evolution for one step of a quantum walk is given by a unitary operator
\begin{equation}
E=(\proj{0}\otimes T_0+\proj{1}\otimes T_1)(U\otimes I)
\end{equation}
\end{definition}
The intuitive picture of the quantum walk in 1D is a particle endowed with
an internal degree of freedom (\textsl{chirality}), which may take values 0 (left) and 1 (right), and
whose state is rotated at each step by $U$. Then the particle takes a step to
the left or to the right, depending on the chirality.

We introduce a new model of the quantum Parrondo's game as follows: We
have four  registers $(\bm{C},\bm{D},\bm{X},\bm{O})$, states of which are
described by vectors in Hilbert spaces $\hilb_C,\hilb_D,\hilb_X,\hilb_O$, respectively. The
register $\bm{X}$ stores the amount of capital; $\bm{D}$ is the coin register
for the strategy used; $\bm{C}$ is the chirality register which determines the
strategy we use; $\bm{O}$ is the auxiliary register. The quantum circuit which
processes the data stored in these registers is shown in Fig.~2.
The quantum Parrondo's
game is defined as
\begin{definition}(Quantum Parrondo's game)
We have $\hilb=\hilb_C\otimes\hilb_D\otimes\hilb_X\otimes\hilb_O$ such that:
\begin{enumerate}
\item The Hilbert spaces $\hilb_j=\mathrm{span}\{\ket{k}:k=0,1\}$ for
$j\in\{C,D,O\}$. All operators on $\hilb_j$ will be henceforth written in the
basis $(\ket{0},\ket{1})$, so that $\ket{0}=(1,0)^T,\ket{1}=(0,1)^T$. The
Hilbert space $\hilb_X=\mathrm{span}\{\ket{x}:x\in\mathbb{Z}\}$

\item $U$ is the unitary operator on $\hilb_C$:
\begin{equation}
U=\frac{1}{\sqrt2}\left[
\begin{array}{ll}
1&\rmi\\
\rmi&1
\end{array}
\right].
\end{equation}

\item The operator $X$ is the NOT gate:
\begin{equation}
X=\left[
\begin{array}{ll}
0&1\\
1&0\\
\end{array}\right].
\end{equation}

\item The operator $\ctrled{A}$ is a controlled $SU(2)$ operator (rotation) on
$\hilb_C\otimes\hilb_D$, the operators
$\ctrled{B_0},\ctrled{B_1}$ are controlled $SU(2)$ operators on
$\hilb_D\otimes\hilb_C\otimes\hilb_O$. In both cases, $\hilb_D$ is the target space.
For any $SU(2)$ operator $G$ we use the parametrization
  \begin{equation}
    G(\theta,\alpha,\beta)=
    \left[
    \begin{array}{ll}
      \rme^{\rmi\alpha}\cos\frac{\theta}{2}&\rmi\rme^{\rmi\beta}\sin\frac{\theta}{2}\\
      \rmi\rme^{-\rmi\beta}\sin\frac{\theta}{2}&\rme^{-\rmi\alpha}\cos\frac{\theta}{2}
    \end{array}
    \right],
  \end{equation}
  with $\theta\in[0,\pi],\alpha,\beta\in[-\pi,\pi]$. We define
  \begin{eqnarray}
  \ctrled{A}=\proj{0}\otimes A+\proj{1}\otimes I  \; ;\\
  \ctrled{B_j}=\proj{0}\otimes
  (B_j\otimes\proj{0}+I\otimes\proj{1})+\proj{1}\otimes I\otimes I \; ,
  \end{eqnarray}
for $j\in\{0,1\}$.
\item The gate $\mathrm{MOD}$ is the conditional operator:
\begin{equation}
    \mathrm{MOD}\ket{x}\ket{o}=
    \begin{cases}
        \ket{x}\ket{o}\quad 3\mid x\\
        \ket{x}\ket{o\oplus1}\quad \mathrm{otherwise.}\\
     \end{cases}
\end{equation}
\item The operator $S$ acting on $\hilb_C\otimes\hilb_X$ updates the $\bm{X}$
register (the capital by)
  \begin{equation}
    S = \proj{0}\otimes T_0+\proj{1}\otimes T_1,
  \end{equation}
  where $T_0\ket{x} = \ket{x-1},T_1\ket{x}=\ket{x+1}$.

\item The gate $\mathrm{MOD_{inv}}$ acting on $\hilb_C\otimes\hilb_X\otimes\hilb_O$
($\hilb_O$ is the target) is the conditional operator which resets the register
$\bm{O}$. If the state of the
  $(\bm{C},\bm{X})$ register at the $n$-th step is $(c_n,x_n)$, we have
  $x_{n-1}\equiv x_n-(2c_n-1)\;(\mod3)$. At the $n$-th step, the
  operator $\mathrm{MOD}_\mathrm{inv}$ flips $\ket{o}$ if and only if
  $x_{n-1}\equiv0\;(\mod{3})$.
\end{enumerate}
\end{definition}

\begin{figure}
  \label{fig:2}
  \centering
  \includegraphics[width=.8\textwidth,keepaspectratio=true]{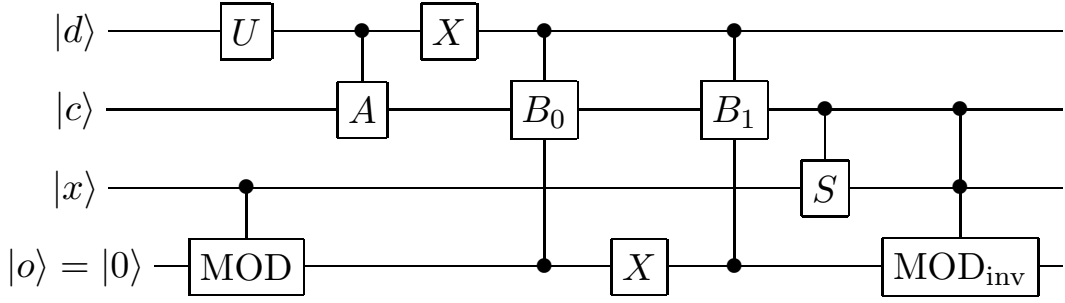}
  \caption{The quantum circuit for the quantum Parrondo's game.}
\end{figure}

The logical circuit shown in Fig.~2 can be simplified to obtain
the circuit presented in Fig.~3.  In this circuit, the operator $W$ acting on
$\hilb_D\otimes\hilb_C\otimes\hilb_X\otimes\hilb_O$ has the form
\begin{eqnarray}
      W=\ket{0}\bra{1}U\otimes\Big[B_0\otimes1\otimes\ket{1}\bra{0}+B_1\otimes1\otimes\ket{0}\bra{1}\Big]
      +\ket{1}\bra{0}U\otimes A\otimes1\otimes X
\end{eqnarray}
and the operators $S,\mathrm{MOD}_\mathrm{inv}$ depend nontrivially only on $\ket{c}$.
\begin{figure}
  \centering
  \label{fig:3}
  \includegraphics[width=.55\textwidth, keepaspectratio=true]{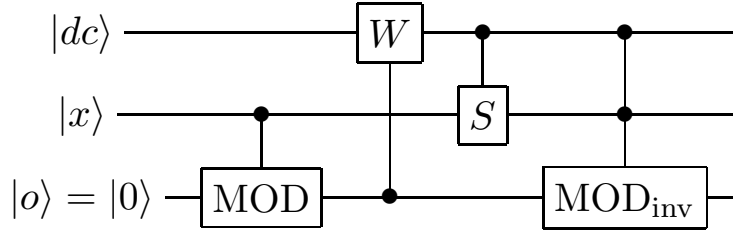}
  \caption{A simplified version of the quantum circuit for the quantum Parrondo's game.}
\end{figure}

We introduce a notation $\hilb_W\equiv\hilb_D\otimes\hilb_C$ and further we express the state of the
whole system using the eigenvectors of the translation operator on $\hilb_X$:
\begin{equation}
    \ket{\phi_k^{j}}=\sum_{\substack{x\in\mathbb{Z} ;  x\equiv j(\mod\,3)}}
    \rme^{\rmi kx}\ket{x},
\end{equation}
for $j\in\{0,1,2\}, k\in[-\pi,\pi]$. It is clear that
$T_0\ket{\phi_k^j}=e^{ik}\ket{\phi_k^{j\ominus1}}$ and $T_1\ket{\phi_k^j}=e^{-ik}\ket{\phi_k^{j\oplus1}}$.
We also set $\ket{\phi_k}=\sum_{j=0}^2\ket{\phi_k^j}$.
The inverse transform is given by an expression
\begin{equation}
    \label{eq:parrondo11}
    \ket{x}=\int_{-\pi}^{\pi}\frac{\rmd k}{2\pi}\rme^{-\rmi kx}\ket{\phi_k}\;.
\end{equation}
The action of $W\cdot\mathrm{MOD}$ on the state
$\ket{\chi}\ket{\phi_k^j}\ket{0}$ gives
\begin{eqnarray}
 \fl \ket{\chi}\ket{\phi_k^0}\ket{0}\mapsto&
  \Big(\proj{01}U\otimes  B_0\otimes1\otimes\proj{10}+\proj{10}U\otimes
  A\otimes1\otimes X\Big)\ket{\chi}\ket{\phi_k^0}\ket{0}  \; ;\\
  \fl\ket{\chi}\ket{\phi_k^{1,2}}\ket{0}\mapsto&
  \Big(\proj{01}U\otimes  B_1\otimes1\otimes\proj{01}+\proj{10}U\otimes
  A\otimes1\otimes X\Big)\ket{\chi}\ket{\phi_k^{1,2}}\ket{1}\; ,
\end{eqnarray}
where $\proj{ab}\equiv\ket{a}\bra{b}$. Application of the operator $S$
and reseting the last register with $\mathrm{MOD}_\mathrm{inv}$ gives
the evolution operator
\begin{equation}
    E=\big(1_{D,C}\otimes1_X\otimes\mathrm{MOD_{inv}}\big)\cdot S\cdot W\cdot\big(1_{D,C}\otimes1_X\otimes\mathrm{MOD}\big)
\end{equation}
whose action on $\ket{\chi}\ket{\phi_k^j}\ket{0}$ is
\begin{eqnarray}
        E\ket{\chi}\ket{\phi_k}\ket{0}=\Big\{(M_{10}+M_{11})\ket{\chi}\ket{\phi_k^0}
        +(M_{10}+M_{01})\ket{\chi}\ket{\phi_k^1}
        +(M_{11}+M_{00})\ket{\chi}\ket{\phi_k^2}\Big\}\ket{0}
\end{eqnarray}
with
\begin{equation}
        M_{jd}=\rme^{s_d\rmi k}(\proj{01}U\otimes\proj{d}B_j+\proj{10}U\otimes\proj{d}A),
\end{equation}
where $j,d\in\{0,1\},s_d=1-2d$.
Multiple application of $E$ on the initial state gives
\begin{equation}
    E^n\ket{\chi}\ket{\phi_k}\ket{0}=\Big(\mu_0^{(n)}\ket{\chi}\ket{\phi_k^1}+\mu_1^{(n)}\ket{\chi}\ket{\phi_k^1}+\mu_2^{(n)}\ket{\chi}\ket{\phi_k^2}\Big)\ket{0}\;.
\end{equation}
The terms $M_j^{(n)}$ are related by the matrix-matrix equation
\begin{equation}
\left[
    \begin{array}{lll}
        \mu_0^{(n+1)}&\\
        \mu_1^{(n+1)}&\\
        \mu_2^{(n+1)}&
    \end{array}\right]=\left[
    \begin{array}{lll}
        0&M_{10}&M_{11}\\
        M_{01}&0&M_{10}\\
        M_{00}&M_{11}&0
    \end{array}\right]
    \cdot\left[
    \begin{array}{lll}
        \mu_0^{(n)}&\\
        \mu_1^{(n)}&\\
        \mu_2^{(n)}&
    \end{array}\right],
\end{equation}
with $\mu_j^{(0)}=1$. The problem can be solved by computing the
eigensystem of this $12\times12$ matrix.
\section{Numerical simulation of quantum Parrondo's game}
In this section we present results of numerical simulations of the quantum Parrondo's game for different initial
states. We assume a coin which is an analogue of the
classical coins $A,B$; namely we consider
$A=G(2(\frac{\pi}{2}-\epsilon),0,0),B_0=G(2(\frac{\pi}{10}-\epsilon),0,0),B_1=G(2(\frac{3}{4}-\epsilon),0,0),\epsilon=\frac{1}{100}$,
We simulate the evolution for up to 1000
steps, counting the expected capital as
\begin{equation}
  c(n)\equiv\sum_{x\in\mathbb{Z}}x\bra{x}\rho_X(n)\ket{x}\; ,
\end{equation}
where
\begin{equation}
  \rho_X(n)\equiv\mathrm{Tr}_{D,C,O}(E^n)\ket{\psi(0)}\bra{\psi(0)}(E^\dagger)^n \, .
\end{equation}
\begin{figure}
  \label{fig:4}
  \centering
  \includegraphics[width=.9\textwidth,keepaspectratio=true]{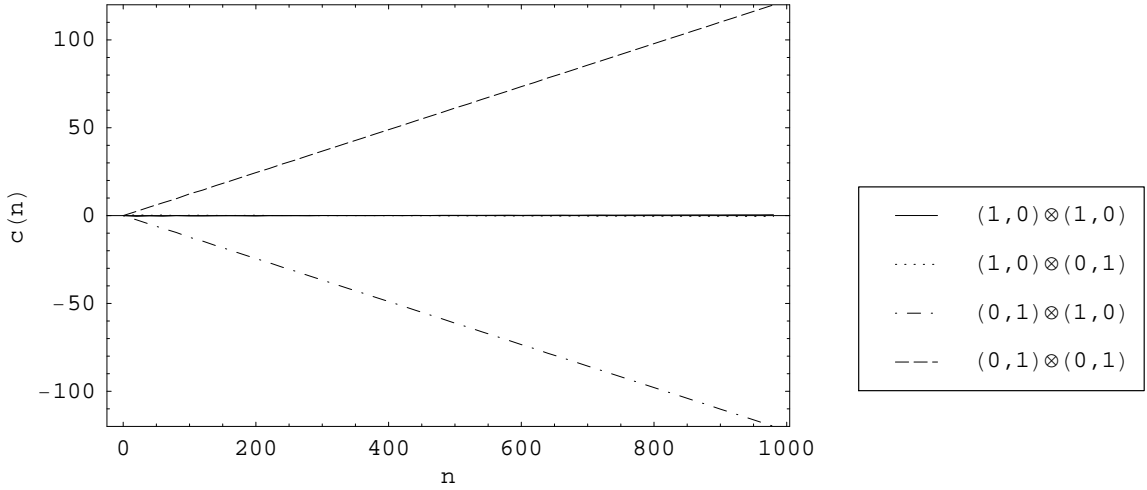}
  \caption{The expected capital $c(n)$ of the quantum Parrondo's game for the
    zero initial capital and different initial states of the registers
    $(\bm{C},\bm{D})$.
  }
\end{figure}
For our purposes we observe four combinations of the basis
states of $\ket{d},\ket{c}$ (see Fig.~4).
We see that the game may be winning, losing or fair, depending on the
initial state of the register $\ket{d}\ket{c}$.
The initial state of $\ket{c}$ determines whether the change in $c(n)$ is
positive or negative (the two being symmetric), while the initial state of $\ket{d}$
determines the size of this change.
 The rate of losing/gaining
the capital is much bigger than for the corresponding classical Parrondo's game with
random switching of the strategies (compare with Fig.~1).

The variance of the expected capital reads
\begin{equation}
 v(n) \equiv \sum_{x\in\mathbb{Z}} x^2\bra{x}\rho_X(n)\ket{x}\; .
\end{equation}
It is easy to see that $v(n)$ does not depend on the initial state of the
register $\bm{C}$, since the evolution is symmetric with respect to the
exchange of directions. However, it does depend on the initial state of
register $\bm{D}$, since this register determines the overall strategy.
The numerical value of $v(n)$ is shown on Fig.~5.
\begin{figure}
  \label{fig:5}
  \centering
  \includegraphics[width=.9\textwidth,keepaspectratio=true]{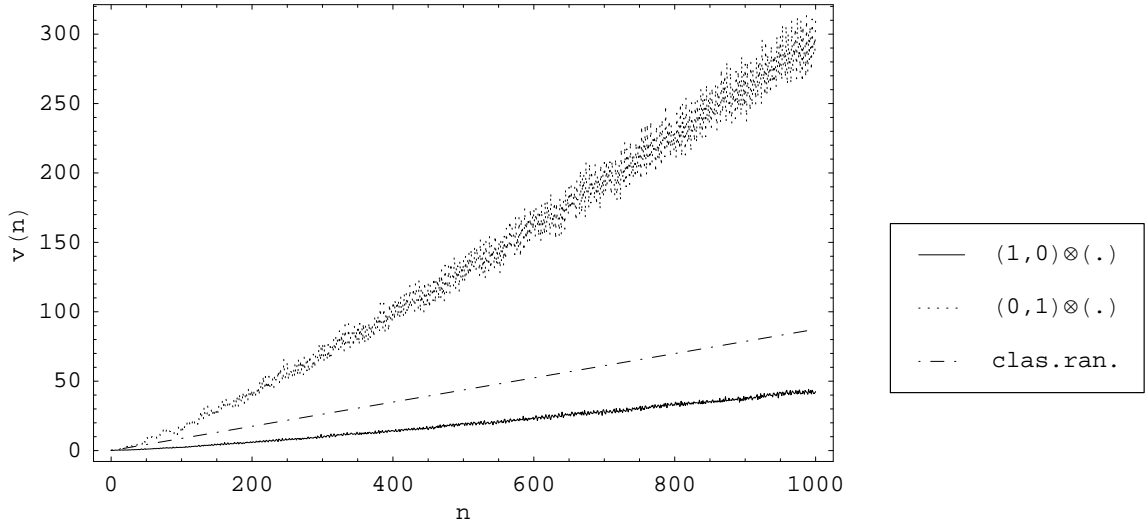}
  \caption{The variance $v(n)$ of the expected capital of the quantum Parrondo's
  game  for the zero initial capital and different initial states of
  register $\bm{D}$. The variance does not depend on the initial state of register $\bm{C}$.
  For comparison purposes we also simulate the variance for classical Parrondo's game with random choice
    of strategies.}
\end{figure}

\section{Random choice of strategies with CP-map}
In  Ref.~\cite{Mey99a} the author considered the difference between quantum
strategies,  and the \textsl{mixed  quantum} strategies. In these mixed strategies one
applies different unitary operators on the qubit with certain
probabilities. It is
probably a better analogue of a random sequence of classical
strategies $A,B$ to consider quantum evolution, where the
application of the operators $A,B_0,B_1$ depends on {\it a priori}
probabilities rather than on a state of the register $\ket{d}$. For
our purposes, we discard the register $\ket{d}$ and the state of the
game is described by a density operator
\begin{equation}
  \rho = \sum_{x,y\in\mathbb{Z}}\rho_{xy}\otimes\ket{x}\bra{y} \; .
\end{equation}
We do not need to consider the state of the register
$\ket{o}$, as any ``garbage'' information which is written into it is
discarded by the operator $\mathrm{MOD}_\mathrm{inv}$. We need to reset the
register $\ket{o}$ so that the projection $1\otimes1\otimes\proj{x}\otimes1$ of
the state vector $\ket{\psi}$ will effectively be from the subspace
$\hilb_D\otimes\hilb_C$.

One step of the
evolution of $\rho$
is described by the CP-map $\mathcal{E}$ such that (we omit the
action of $\mathrm{MOD},\mathrm{MOD}_\mathrm{inv}$)
\begin{equation}
\label{eq:parrondo12}
  \fl\mathcal{E}[\rho]  =  \sum_{j,k\in\{0,1\}}(\proj{j}\otimes
  T_j)\Big[\sum_{x,y\in\mathbb{Z}}\frac{1}{2}\big(A\rho_{xy}A^\dagger+B_{01}\rho_{xy}B_{01}^\dagger\big)\otimes\ket{x}\bra{y}\Big]\big(\proj{k}\otimes
  T_k\big)^\dagger \; .
\end{equation}
Here $B_{01}$ depends on a state of the register $\ket{o}$ in the usual
way. In this dynamics respective operators are applied with probabilities equal to $\frac{1}{2}$.

\section{Numerical simulation of mixed Parrondo's game}
We have simulated the evolution of the mixed Parrondo's game for
different initial states of the register $\ket{c}$ and zero initial
capital. The results (see Fig.~6) show that
dynamics is different from both the random Parrondo's game and the quantum
Parrondo's game in that the capital converges to a stationary value,
which is either positive or negative, depending on the initial state
of $\ket{c}$.
\begin{figure}
  \label{fig:6}
  \centering
  \includegraphics[width=.8\textwidth,keepaspectratio=true]{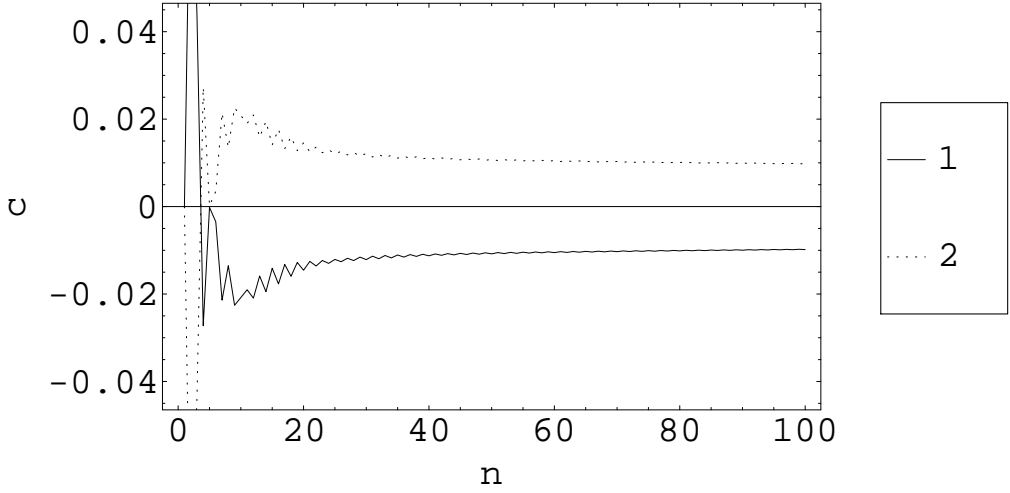}
  \caption{The capital of the mixed Parrondo's game for zero initial
    capital and the initial state of the $\bm{C}$ register being
  $\ket{0}$ (1), $\ket{1}$ (2).}
 \end{figure}
The expected capital resulting from the evolution given by Eq.~(\ref{eq:parrondo12})
depends on the initial state in a symmetric way. To see this, let us consider the evolution
\begin{equation}
    \mathcal{E}[(X\otimes1)\rho(X^\dagger\otimes1)] \; ,
\end{equation}
where $X$ is the swap operator on $\hilb_C$. It is obvious that $X$
commutes with $A$ and $B_0,B_1$, and $(\proj{L}\otimes T_L+\proj{R}\otimes
T_R)(X\otimes1)=(X\otimes1)(\proj{R}\otimes T_L+\proj{L}\otimes T_R)$. Hence,
$n$ steps of  the evolution with the swapped state give
\begin{equation}
    \mathcal{E}^n[(X\otimes1)\rho(X^\dagger\otimes1)] = (X\otimes
    Y)\mathcal{E}^n[\rho](X\otimes Y)^\dagger \, ,
\end{equation}
where $Y\ket{x}=\ket{-x}$. The symmetry in the expected capital with respect to initial
states $\proj{L}\otimes\proj{0}$ and $\proj{R}\otimes\proj{0}$ as seen in
Fig.~6 immediately follows.
The variance of the expected capital resulting from the mixed Parrondo's game is
shown in Fig.~7. Since the capital depends symmetrically on the initial state of
$\bm{C}$, the variance is independent of it.
\begin{figure}
  \label{fig:7}
  \centering
  \includegraphics[width=.7\textwidth,keepaspectratio=true]{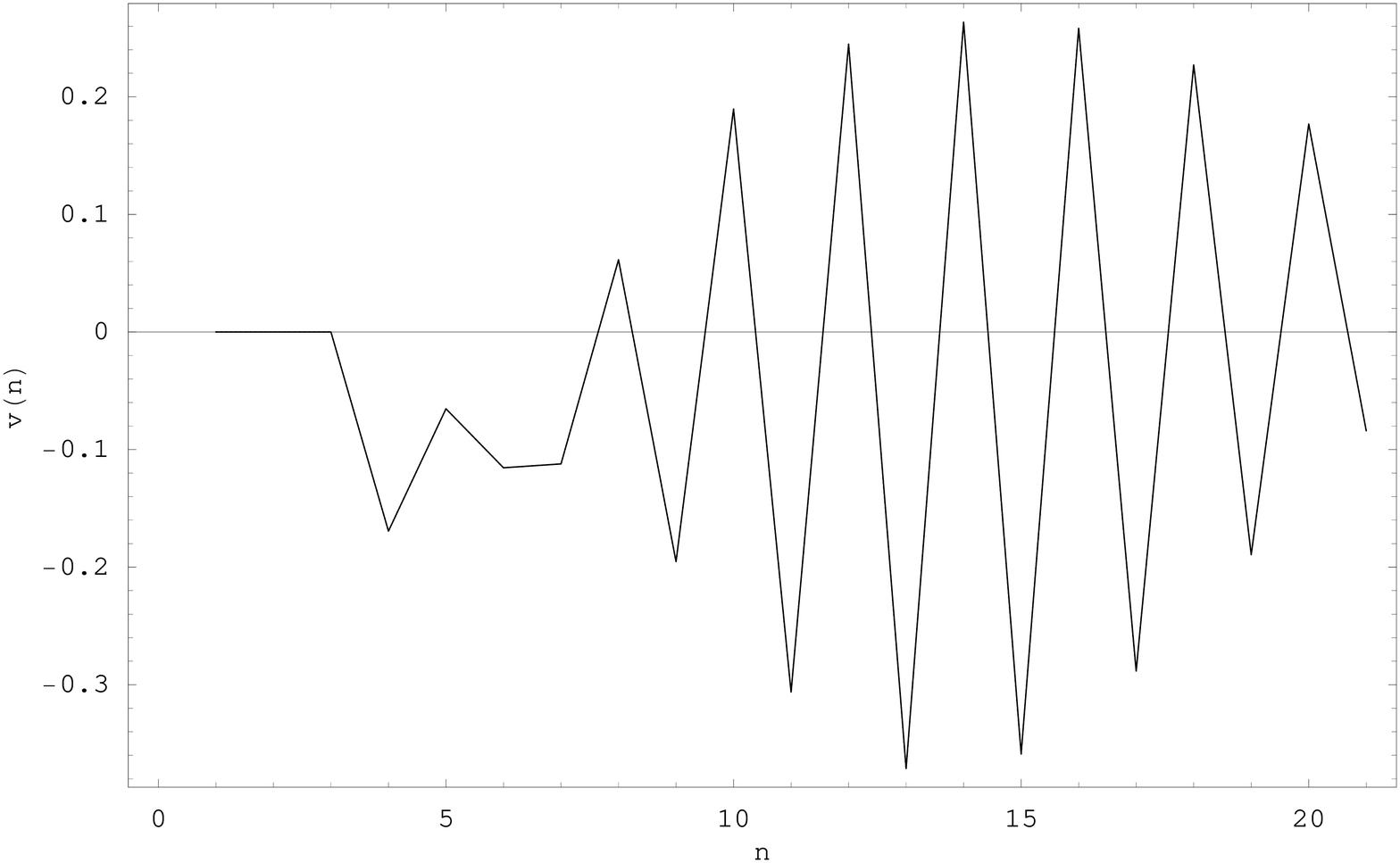}
  \caption{The variance of the expected capital of the mixed Parrondo's game for
  zero initial capital and the initial state of the $\bm{C}$ register being
  either $\ket{0}$ or $\ket{1}$.}
 \end{figure}

\section{Conclusions: Connection between Parrondo's games}
The natural question arises: what is the connection between the
three versions of Parrondo's game we have considered in this paper?
The quantum Parrondo's game may be transformed into both the classical
Parrondo's game with quasi-random (memory dependent) strategies, and
the mixed Parrondo's game. To see this, let us consider that in the quantum Parrondo's game we
measure the register $\bm{D}$ at  each step (just after the application
of the operator $U$).  If the initial state
of $\bm{D}$ is either $\ket{0}$ or $\ket{1}$, the operator $U$
prepares equally weighed superposition of states
$\ket{0},\ket{1}$. Measurement of  the register $\bm{D}$ gives a uniform
probability distribution over $\ket{0},\ket{1}$, hence the rest of the dynamics
corresponds to random choice of strategies of $A,B$ and we obtain the
mixed Parrondo's game (compare Fig.~6 and Fig.~8).
\begin{figure}[htb]
  \label{fig:8}
  \centering
  \includegraphics[width=.8\textwidth,keepaspectratio=true]{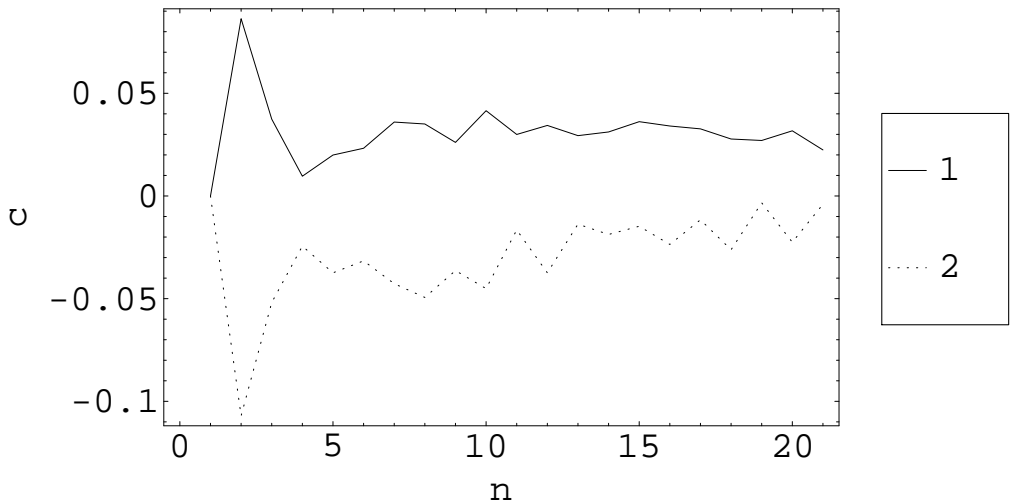}
  \caption{The evolution of the capital which arises when we make a measurement
  of the register $\bm{C}$ immediately after we apply the operator $U$, averaged
  over 5000 samples. The initial state of $\bm{D}$ is $\ket{0}$ and the initial
  state of $\bm{C}$ is $\ket{0}$ (1),$\ket{1}$ (2).}
\end{figure}

Moreover,  let us consider that we also measure the
register $\bm{C}$ at each step (after the action of
$A,B_0,B_1$). Then the state of $\ket{c}$ collapses onto
$\ket{0},\ket{1}$ (with the biased probability) and the state of $\ket{x}$
is changed to the orthogonal state $\ket{x\pm1}$. However, this
evolution differs from the classical Parrondo's game in that the bias
of the coins depends on the outcome of the last measurement. To see
this, let us consider that the initial state of $\bm{C}$ is $\ket{0}$ and in the
first step we apply $A$. Then the new state of $\bm{C}$ is
$[\sin\epsilon\ket{0}+\rmi\cos\epsilon\ket{1}]$ and the measurement
on $\ket{c}$ gives $\ket{0},\ket{1}$ with respective probabilities. If
at the next step we happen to apply $A$ again, the new state of
$\bm{C}$ will be either $[\sin\epsilon\ket{0}+\rmi\cos\epsilon\ket{1}]$
(if the last measurement gave $\ket{0}$) or
$[\sin\epsilon\ket{1}+\rmi\cos\epsilon\ket{0}]$ otherwise. We see that the
bias to measure $\ket{0},\ket{1}$ changed (in classical terms, the new
coin toss is more likely to win, if the last coin toss was losing, and
vice versa).

In this paper we have shown how we can implement the Parrondo's game with random switching of
strategies using quantum formalism, and what is the difference between the
``randomness'' in the sense of quantum walks and the true randomness implemented
via CP-maps. The first case leads to strictly positive or negative gain in the
capital, or even to zero outcome, depending on the initial state of the coin
registers. The second case may also lead to the positive or negative gain;
however the capital converges to a fixed value.
The measurement of a selected register may reduce the quantum Parrondo's game to the
mixed Parrondo's game, and hence suppress the winning ratio of the game.

Finally, we note that there exist other versions of the quantum Parrondo's game.
Specifically,  in Ref.~\cite{FAJ03a} the authors discussed how the paradox arises when
coin tosses depend on the states of the coins at the previous steps. Cooperative Parrondo's game are also of interest.
The problem of coins with memories and other modifications of Parrondo's game as well as
physical realization of the game via quantum walks will be presented elsewhere.

\textit{Acknowledgments}: We thank Mark Hillery and Jason Twamley for helpful
discussions.
This research was supported in part by the European Union  projects QAP,
 CONQUEST, by the INTAS project 04-77-7289,  by
the Slovak Academy of Sciences via the project CE-PI/2/2005, and by the
project APVT-99-012304. The project was also partially funded by Polish Ministry of Science and Higher
 Education grant number N519 012 31/1957.

%\bigskip


\begin{thebibliography}{12}

\bibitem{NM44a}
J. von Neumann and O. Morgenstern,
{\itshape Theory of games and economic behavior}
(Princeton University Press, Princeton, 1953).

\bibitem{EWL99a}
J. Eisert, M. Wilkens M, and M. Lewenstein,
Phys. Rev. Lett. {\bf 83}, 3077 (1999).

\bibitem{Nas51a}
J. Nash,
The Annals of Mathematics {\bf 54}, 286 (1951).

\bibitem{Mey99a}
D. A. Meyer,
Phys. Rev. Lett. {\bf 82}, 1052 (1999).


\bibitem{Sanders2003}
B. C. Sanders, S. D. Bartlett,  B. Tregenna, and P. L. Knight,
Phys. Rev. A {\bf 67}, 042305  (2003).


\bibitem{Knight2003a}
P. L. Knight, E. Roldan, and J. E. Sipe,
Phys. Rev. A {\bf 68}, 020301 (2003).


\bibitem{Knight2003b}
P. L. Knight, E. Roldan, and J. E. Sipe,
Optics Communications  {\bf 227}, 147 (2003).


\bibitem{Knight2004}
P. L. Knight, E. Roldan, and J. E. Sipe,
J. Mod. Opt.  {\bf 51}, 1761 (2004).


\bibitem{Carneiro2005}
I. Carneiro, M. Loo, X. B. Xu, M. Girerd, V. Kendon, and P. L. Knight,
New Journal of Physics {\bf 8},  156 (2005).


\bibitem{HA99a}
G. P. Harmer and D. Abbott,
Nature {\bf 402}, 864 (1999).

\bibitem{PHA00a}
J. M. R. Parrondo, G. P. Harmer, and D. Abbott,
Phys. Rev. Lett. {\bf 85}, 5226 (2000).

\bibitem{PD04a}
J. Parrondo and L. Din{\'\i}s,
Contemporary Physics {\bf 45}, 147 (2004).

\bibitem{MB02a}
D. A. Meyer and H. Blumer,
Journal of Statistical Physics {\bf 107}, 225  (2002).

\bibitem{TAM03a}
R. Toral, P. Amengual, and S. Mangioni,
Physica A {\bf 327}, 105 (2003).

\bibitem{MG05a}
J. A. Miszczak and P. Gawron,
Fluctuation and Noise Letters  {\bf 5},  (2005).

\bibitem{FNA02a}
A. P. Flitney, J. Ng, and D.~Abbott,
Physica A {\bf 314}, 35 (2002).

\bibitem{ADZ93a}
Y. Aharonov, L. Davidovich, and N. Zagury,
Phys. Rev. A {\bf 48}, 1687 (1993).

\bibitem{RSS+04a}
A. Romanelli, A. C. Sicardi Schifino, R. Siri, G. Abal,  A. Auyuanet, and
R. Donangelo,
Physica A {\bf 338}, 395 (2004).

\bibitem{FAJ03a}
A. P. Flitney, D. Abbott, and N. F. Johnson,
Quantum random walks with history dependence.
Available online at: http://arxiv.org/abs/quant-ph/0311009v1.


\end{thebibliography}
\end{document}